\documentclass[]{spie}  

\usepackage{amsmath,amsfonts,amssymb}
\usepackage{graphicx}
\usepackage[colorlinks=true, allcolors=blue]{hyperref}
\usepackage{astro_bib_macro}

\usepackage{booktabs}

\usepackage{listings}

\usepackage{enumitem}
\newlist{arrowlist}{itemize}{1}
\setlist[arrowlist]{label=$\Rightarrow$}

\title{Connecting the astronomical testbed community - the CAOTIC project: Optimized teaching methods for software version control concepts}

\author{Iva Laginja\supit{a}, Pablo Robles\supit{b,c}, Kevin Barjot\supit{a}, Lucie Leboulleux\supit{d}, Rebecca Jensen-Clem\supit{e}, Keira J. Brooks\supit{f}, Christopher Moriarty\supit{g}
\skiplinehalf
\supit{a} LESIA, Observatoire de Paris, Universit\'{e} PSL, Sorbonne Universit\'{e}, Universit\'{e} Paris Cit\'{e}, CNRS, 5 place Jules Janssen, 92195 Meudon, France\\
\supit{b} DOTA, ONERA, Universit\'e Paris Saclay, F-92322 Ch\^{a}tillon, France\\
\supit{c} Aix Marseille Universit\'{e}, CNRS, LAM, UMR 7326, 13388 Marseille, France\\
\supit{d} Univ. Grenoble Alpes, CNRS, IPAG, 38000 Grenoble, France\\
\supit{e} University of California Santa Cruz, Santa Cruz, CA 95064, USA\\
\supit{f} Laboratory for Atmospheric and Space Physics, Boulder, CO, USA\\
\supit{g} Center for Astrophysics, Harvard \& Smithsonian, Cambridge, MA 02138, USA
}

\authorinfo{Further author information, send correspondence to Iva Laginja: E-mail: iva.laginja@obspm.fr}

\begin{document} 
\maketitle

\begin{abstract}
Laboratory testbeds are an integral part of conducting research and developing technology for high-contrast imaging and extreme adaptive optics. There are a number of laboratory groups around the world that use and develop resources that are imminently required for their operations, such as software and hardware controls. The CAOTIC (Community of Adaptive OpTics and hIgh Contrast testbeds) project is aimed to be a platform for this community to connect, share information, and exchange resources in order to conduct more efficient research in astronomical instrumentation, while also encouraging best practices and strengthening cross-team connections. In these proceedings, we present the goals of the CAOTIC project, our new website, and we focus in particular on a new approach to teaching version control to scientists, which is a cornerstone of successful collaborations in astronomical instrumentation.
\end{abstract}


\section{Introduction}
\label{sec:introduction}

Laboratory testbeds for high-contrast imaging (HCI) and extreme adaptive optics (AO) systems are indispensable venues for technology development in a field that relies on the largest-aperture ground and space-based telescopes. Ever more ambitious requirements for astronomical observations, such as higher sensitivity, finer resolution, and imaging at deeper contrasts are driving the development of new and improved techniques in the field of astronomical instrumentation\cite{astro2020}. Specifically, the research in HCI and AO is very technically oriented and laboratory testbeds are a crucial component in every project, providing facilities from proof-of-concept realizations over testing grounds, to full system demonstrations\cite{Mazoyer2019HighContrastTestbedsFuture}. The operation of such testbeds requires expertise in all scientific and technical areas, including hardware controls, software engineering, data storage and management, as well as processes to thread all of these components into an overall project that runs smoothly and robustly.

To date, there are more than a dozen of such testbeds at various institutions around the world, focusing on one or several of the following topics: exoplanet imaging, coronagraphy, wavefront sensing and control, adaptive optics, image processing, data analysis and component development (e.g., detectors, mirrors). Every research group is recording remarkable results in their respective project regions, however, the communication and exchange between the groups is inherently limited to published papers and proceedings, conference talks, and sparse email contact. We have identified the potential for easily accessible testbed information and the ease of communication across the community that aims to eliminate the need for each team to ``reinvent the wheel'' when implementing hardware and software solutions, as well as facilitate cross-testbed learning. Since all of these facilities use a finite number of well-known hardware components and expand well-established optical algorithms, our aim is to provide a platform for exchange. In this way, we hope to standardize certain approaches taken in the implementation and maintenance of the testbeds and to accelerate the research findings coming out of this community.

The CAOTIC (Community of Adaptive OpTics and hIgh Contrast testbeds) project provides a platform to leverage this potential and it is currently represented by a website. Through submissions from the community, we have collected technical specifications and information about their operations from a little over a dozen testbeds with the goal of providing a top-level overview of the field. By promoting the sharing of resources, the development and use of open-source software, and higher visibility for junior people - students, postdocs, and young professionals - in the field, our target is to strengthen the ways in which we build networks, spread knowledge and give access to information.

In particular, the first point in the actionable content we identified in the context of the CAOTIC project is to rethink what standards we as a community want to be able to rely on for the purpose of software management. As a core component of many testbed projects, software development is a technical topic that is too often left in the hands of a purely self-taught workforce in this field, namely astronomers creating full software environments and infrastructures. While the researchers involved in creating and running these optical testbeds are undoubtedly the experts who decide on the project goals and their execution, there is a stark lack of software development skills within this demographic. Efficiently addressing certain questions, like creating a new control architecture, rewriting code in a different language, introducing a level of abstraction or setting up a continuous integration framework, can sometimes only be done through the hiring of one or more software engineers by training. Other needs though, like the implementation of new algorithms, the integration of individual drivers and overall maintenance of an already well-designed project can easily be met by the scientific staff. However, to keep the interactions between the different software needs and implementations smooth, it is necessary to find a way to consolidate the software management process between the individual contributors.

In particular, efficient collaboration on code and its versioning and safe-guarding through backups is one of the most important aspects in this process. In most software projects outside of academia, this need is addressed by using version control systems (VCS). While this concept is not unknown to the scientific community, it is often ignored when setting up a testbed project or more generally research projects involving software development. One of the reasons for this is the lack of understanding of the goals and workings of VCS in the wider astronomy community, and the inherent lack of training opportunities for this particular topic. This is why we decided to address this issue as the first main goal of the CAOTIC project.

In these communications, we start in Sec.~\ref{sec:testbeds-and-labs} by introducing a broad overview of the general work of astronomical testbeds and laboratories before motivating the need to move to standardly using VCS. In Sec.~\ref{sec:caotic-project}, we present the overall goals of the CAOTIC project, its current status, impact thus far, and plans for the second half of 2022 and beyond. In Sec.~\ref{sec:git-github-teaching}, we highlight the goals and methods of a new approach to teaching version control for software development in research projects and present the impact of the version control course series tailored to scientists that took place in the first halves of 2021 and 2022. Finally, in Sec.~\ref{sec:summary}, we conclude our work and give an outlook for the future of these activities.

\section{Astronomical testbeds and laboratories}
\label{sec:testbeds-and-labs}

Astronomical instrumentation is a wide field of research with many scientific applications. Since most objects of astronomical research cannot be captured and brought to, or replicated in a laboratory, most instrumental applications focus on developing the technology to conduct observations of faraway objects with the goal to make new discoveries and confirm theoretical models. The building and testing of these instruments is mostly left to engineering teams, with some input from the scientists involved in the particular project, but the role of the latter usually becomes dominant only once an instrument starts its on-sky operations.

There are certain applications though where the development and improvement of optical instruments represents the concrete scientific work itself. This is in particular true for the field of direct imaging, where a project can consist of designing and building a testbed which is consequently used to test new instrumental methods or algorithms. In this case, the experimental results themselves are the end goal in order to set the path for consecutive testbeds or future on-sky instruments. The technologies that shape the field of direct imaging are coronagraphy; wavefront sensing and control (WFS\&C) including hardware and algorithms for both wavefront sensing (WFS) and wavefront control (WFC); focal-plane WFS; adaptive optics and predictive control; as well as post-processing methods.

Astronomical testbeds are an integral part of developing these technologies and can serve various purposes on the way to developing fully mature direct imaging instruments:
\begin{itemize}
    \item Component-level development and testing, e.g. new coronagraph masks or WFC algorithms.
    \item Systems development, e.g. the architecture of AO systems and interplay between different starlight suppression components, sensors and controllers.
    \item Laboratory and on-sky demonstrations of fully integrated systems, and related trade-off studies.
\end{itemize}
While each project is pursuing its own goals, the tools and methods in doing so have become more and more common. This can be the same hardware equipment, for example the same camera models or laser sources, but this is especially true for critical components like deformable mirrors (DM) - there exists a finite number of both continuous face-sheet and segmented DMs, from a limited number of manufacturers, so different projects are bound to be confronted with the same hurdles when integrating them onto a testbed. This often encompasses general hardware work and organization, for example cable management or how to establish a remote connection to laboratory computers. But this can also manifest itself on the software side of a project, where the same task (e.g., writing a controller for a DM or camera) keeps seeing repeated reimplementation by different teams. While there is certainly some need for customization in these solutions, there is no need to redo all parts of the infrastructure from scratch.

Sharing software through open-science approaches is not a new concept and especially the last decade has seen a significant increase in scientific software packages being distributed freely to peers. In particular the use of GitHub\cite{github}, a cloud-based software hosting service with a plethora of tools for software development base on the open-source VCS git\cite{git}, has become the go-to solution for shared resources within the scientific community\cite{Perkel2016}. This includes astronomy, where also leading space agencies like NASA, ESA and CSA (US, European and Canadian space agencies, respectively) have embraced the open-source approach for collaboration\cite{Numrich2022}. There has been significant work put into various initiatives supporting this strategy, like the OpenAstronomy project\cite{openastronomy} and more recently, NASA's Transform to Open Science (TOPS)\cite{nasa_tops}. The need to support this path forward has been identified as critical in order to fully exploit the opportunities from shared resources in the future\cite{Tollerud2019Sustaining}. In the case of astronomical instrumentation and optics, some open-source packages have established themselves as a viable resource for optical propagations and simulations, like Poppy\cite{Perrin2016POPPYPhysicalOptics}, PROPER\cite{Krist2007Proper} or HCIPy\cite{Por2018HighContrastImaging}. Equally, some projects dealing with software infrastructures for hardware control have been made available to the community, for example CACAO\cite{cacao}, catkit\cite{Noss2022catkit} and milk\cite{milk}.

Independently of the tools and implementation of open-source projects within astronomy, it is clear that the work force that is anticipated to create and use them needs to have the appropriate skills\cite{Norman2019Growing}. This includes people working in instrumentation and in particular on astronomical testbeds, where the need for efficient software management is immediately apparent. The problems faced by such teams includes the fact that team members come and go - postdocs and students make up a large fraction of the work force, but their time within a team is usually limited to 2--4 years, while the project itself is usually designed to run for longer. In the beginning of their appointment they need to learn the specific tools used on their particular team, and towards the end they need to perform a transfer of knowledge before moving on. A team that minimizes the on-boarding time and integrates new ideas into the overall project as they grow instead of just before the departure of a team member is able to shift the focus from these procedural tasks more to the scientific results themselves.

One of the most involved processes to get acquainted with on a new team is software management. How to integrate one's own software contributions into the overall laboratory infrastructure and how to deploy it on a testbed in a reproducible and robust manner often relies on case-by-case examples that are not uniform across the project, let alone across different laboratories. The big asset for teams here is the use of VCS. Version control is a concept well known to the more tech-savvy individuals in astronomy, but the opportunities it offers are still being widely ignored by the wider community. Since VCS is conceptually completely independent from any chosen software implementation or its distribution, it harbors a big potential for standardization without impeding the individual character of each individual project. It is thus that the usage of version control is one of the main tenets of the CAOTIC project. We believe that it can significantly contribute to the success of a testbed project and its scientific results and that it can improve the exchange of knowledge and skills between different projects, thus advancing the field of astronomical instrumentation as a whole.

\section{The CAOTIC project}
\label{sec:caotic-project}


The project is currently centered around a website that aims to be a low-maintenance platform where interested members of the community can contribute to and organize relevant resources. Initially, this was realized with a Google website which went online in 2017. Ultimately, this did not fall in line with one of the main goals of the project -- to provide an easy and fast exchange of information between different instrumentation groups -- since only the page admins were able to change its content. Thus, the website was migrated to GitHub in October of 2018\footnote{Website URL: \url{https://highconaotools.github.io/}\\GitHub repository: \url{https://github.com/highconaotools/highconaotools.github.io}}. Hosting such a project on GitHub has the advantage that anyone can draft additions and changes to the website and then request their integration. This solution would combine the goal of providing a community platform with the goal to promote software best practices in research.
    \begin{figure}
    \centering
   \includegraphics[width = \textwidth]{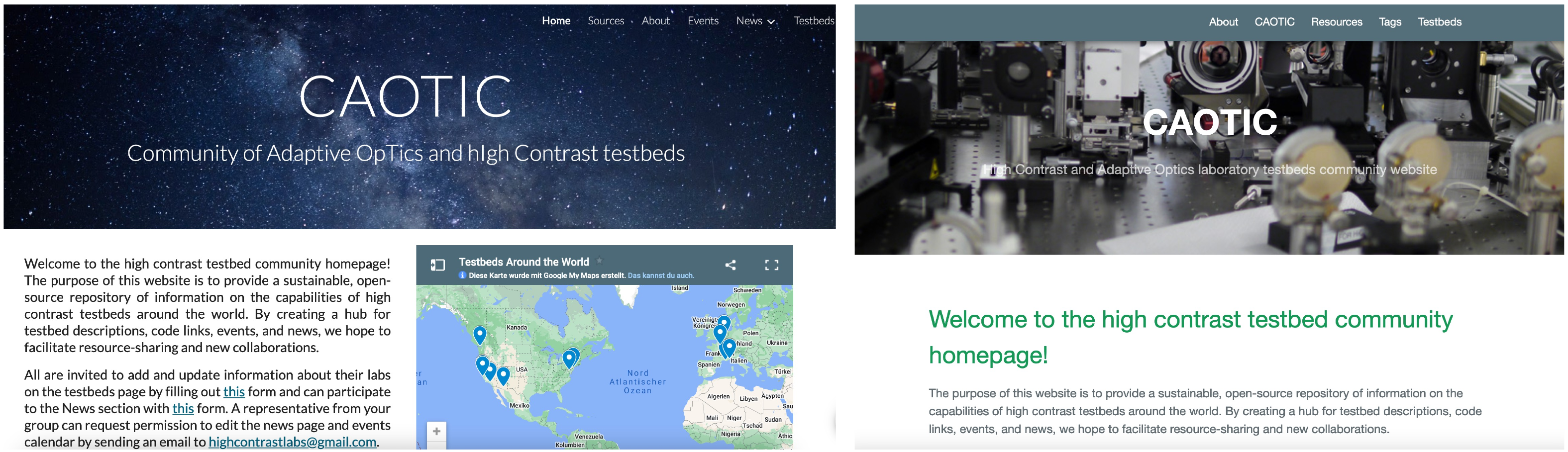}
   \caption[Website] 
   {\label{fig:website} 
    \textit{Left:} Initial website for the CAOTIC project, created with Google sites. \textit{Right:} Current GitHub-hosted project website. Hosting it on GitHub makes the website more accessible to contributions.}
   \end{figure}

The core of this website contains a table listing participating testbeds and their basic information like location, science goals, key hardware components, as well as involved team members and their contact info. This data is complemented by a list of software resources and, in the future, will be extended with relevant talks, literature, courses and events. Since, as stated above, the main reason for hosting the CAOTIC website on a GitHub repository is to facilitate the contribution of new content or changing existing content as easy as possible for anyone within this community, the project provides pre-made templates for new contributions which can be published by means of a pull-request on GitHub, after review by the project owners. This requires a basic understanding of version control with git, as well as the GitHub platform. Since one of the dedicated goals of the CAOTIC project is to promote best practices in software development, which includes the use of version control, the first big action item within the scope of the project was the launch of a series of workshops about using version control with git for the purpose of academic research, which is described in detail in the next section.

\section{Adapted teaching: Git and GitHub for scientists}
\label{sec:git-github-teaching}

In the following section, we present our take on why using version control tools is indispensable for laboratory teams, how to adapt them to the specific needs of astronomers, how we optimized the teaching of such tools and the feedback we obtained in the process.

\subsection{Establishing version control as a standard tool in research}
\label{subsec:version-control-standard}

There are certain tools that facilitate scientific work and thus also the work that a lab creates that we tend not to question anymore. One example is the use of the Smithsonian/NASA Astrophysics Data System (ADS), also known as the ADS Abstract Service\cite{Kurtz1993} for bibliographic search. It has greatly developed since its launch in 1993 and rarely would anybody try to use any different search engine to do bibliographical searches in the domains of physics and astronomy (although other tools exist and have their place, for example Google Scholar). Similarly, a vast majority of manuscripts in astronomy today are prepared using the markup language ``\LaTeX''\cite{Rowley2001}\footnote{\url{https://www.latex-project.org//}}, for many reasons. This includes but is not limited to\cite{Sinclair2018}: consistent typesetting and formatting across a document or several documents, the simplicity to write mathematical expressions, bibliography management and easy sharing between collaborators with cloud-based tools like Overleaf\footnote{\url{https://www.overleaf.com/}}. Not everybody uses it, and it is not being used for every manuscript ever written; however, the crucial point is that almost every single astronomer has used it at least once in their life or participated in a project that required them to use it. While preparing manuscripts in \LaTeX~might not be the best option in all cases, it is acknowledged that it has a firm place in a researcher's tool box, so much so that many institutions and universities offer classes teaching their students and faculty how to use it. 

Writing various types of papers, proposals and reports undeniably makes up a huge fraction of research, but as it turns out, so does writing software. Especially when working in an astronomical laboratory, software engineering represents a continuous thread through the many aspects of designing, building, and operating an optical testbed. A lot of effort is going into writing code to control the mechanical components of the testbed, synchronize them to perform experiments, perform high-fidelity optical simulations, implement a variety of algorithms and analyze the resulting data, a very slim subset of which we mention in Sec.~\ref{sec:testbeds-and-labs}.
Such software tools can be written in many different programming languages that have become more or less popular in the scientific community over the years: from Fortran, to IDL, Mathematica, Matlab, C++ and Python to the more recently developed Julia, to only name a few. This paper does not intend to be a discussion of the different trade-offs between these languages, nor a promotion of any language in particular. Instead, we strongly adhere to the claim that \textbf{every single research project, no matter the programming language it uses, benefits from using version control}. Further, we insist that the current most used way of teaching version control is outdated and unadapted for the needs of a researcher, which we come back to in Sec.~\ref{subsec:version-control-for-astronomers}. There are three main reasons we believe that version control and its associated technologies should be a skill acquired by everybody in the broad astronomical workforce:
\begin{enumerate}
  \item The version control aspect itself
  \item The intrinsic benefits of backing up one's work
  \item The enabling of collaborating with other researchers more efficiently.
\end{enumerate}

\begin{figure}
    \centering
   \includegraphics[width = 0.8\textwidth]{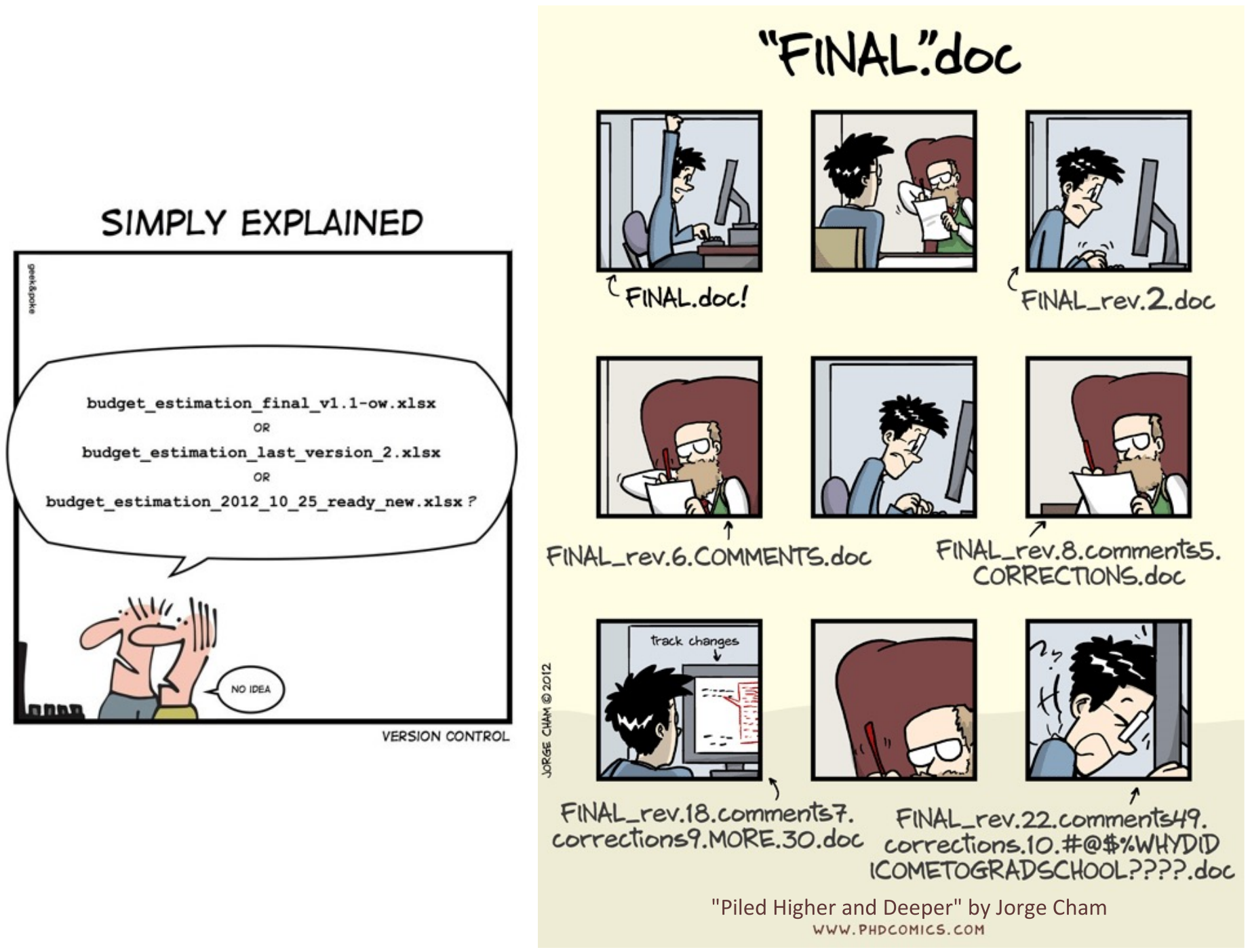}
   \caption[Version control] 
   {\label{fig:version-control} 
    The main, but not only motivation for using version control: the version control aspect itself. \textit{Left:} Comic by Simply Explained\cite{version-control_simply-explained}. \textit{Right:} Comic from ``Piled Higher and Deeper'' by Jorge Cham\cite{version-control_phd-comics}.}
   \end{figure}

The first point in the above list is comically, but also very realistically depicted in the two illustrations in Fig.~\ref{fig:version-control}. In most lines of work, but especially in very explorative fields like science, there is often the desire or need to be able to roll back to a previous version of a product, or in our example, code. The simple solution at first seems to be a straight ``copy-paste and rename'' like in the given illustrations, but it is very easy to loose track of the properties of each given version when versioning is handled this way, especially when coming back to a particular project after weeks or months.
   
\begin{figure}
    \centering
   \includegraphics[width = 0.9\textwidth]{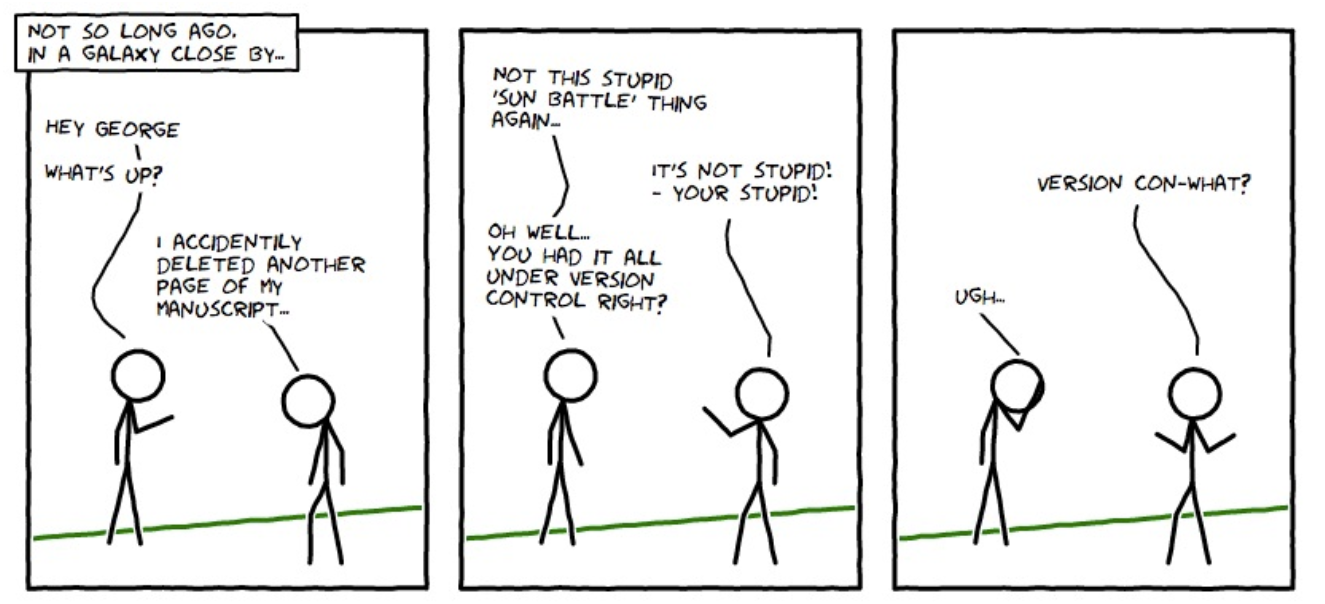}
   \caption[Backing up your work] 
   {\label{fig:backing-up-work} 
    Losing your work due to insufficient backups is one of the worst nightmares of a researcher. Image \copyright~Simon Mutch's Version Control Tutorial\cite{galaxy}, licensed under a Creative Commons Attribution-ShareAlike 3.0 Unported License, no changes applied.}
   \end{figure}

The second point in the above list can almost be considered a useful benefit when using a distributed version control tool like git\cite{git}. It means that there will always be at least one full copy of the project saved on a remote server, preventing any work from being lost in case of the failure of a researcher's work machine. In most cases, this feature comes completely for free when working with version control including a remote repository, which is a dedicated location to save a project. Accidents happen and having a solid backup system in place for one's work can be a life-saver (see also Fig.~\ref{fig:backing-up-work}).

The third point in the above list is enabled by the wide range of collaborative version control tools now available freely to anybody with internet access. By creating a remote repository on a shared-access service like GitHub\cite{github}, Bitbucket\cite{bitbucket} or a personalized GitLab\cite{gitlab} installation, a copy of the version-controlled code base can be made accessible to anybody with an email address and who is familiar with git. By selecting a standardized way of doing this, it becomes very easy to bring new collaborators on board with a new project that requires the development of code. This is true for larger teams who work together on a single project like the controls for a testbed, but it is equally true if as little as two or three people need to use and change the same software project that can be a data reduction pipeline, a simulation tool or any code-based project. Too often are snippets of code still sent around by email, in which case all connection to the previous history of the code is lost. This is sometimes countered by encoding all metadata into the code itself in the form of comments; but rarely is this sufficient to keep up a cohesive and complete history, let alone have a way to synchronize between diverging code bases that have now been completely separated from each other.

There are many more motivations to use version control, but we would like to bring up three more:
\begin{enumerate}
\setcounter{enumi}{3}
  \item The need to always have one working version of the code
  \item Good research practices: repeatability, traceability, open science
  \item The portability to non-academic jobs.
\end{enumerate}
One aspect in the discussion of version-controlled projects that is obvious to any software engineer but often overlooked by scientists is that VCS allow you to always keep one or more functional versions of the code available at all times, while new and potentially buggy features that are still under development can be handled separately, which is reflected in point 4 of the above list. This is very useful in most use-cases like developing a new simulator, where new features are supposed to enhance the functionality and not first break it before debugging your way to the new version. This is also true for code used for data analysis, you might want to start work on coding up a new feature while still run the currently working analysis in the background. And obviously, this is very important when the project in question is the operation of a testbed. While upgrades and enhancements are being coded up and prepared for testing, the testbed can always be used to run experiments without down-times due to untested and buggy code in the main version. (This does not help with fighting the lab gremlins and their vicious, inexplicable and sometimes transient malfunctions of testbed hardware, but you do what you can.)

We wanted to specifically point out an ethical motivation to keep one's software history clean and reproducible in point 5, and that is scientific integrity. Not being able to revert to the version of the code that produced the data and figures for your paper some years ago can pose a significant breach of scientific integrity. Keeping your work traceable and repeatable, possibly even archived by version in a public software archive can solve this problem\cite{open-science-git}.

In point 6 of the above list we want to emphasize the utility of general technical skills to career paths outside of academia. Especially in the field of astronomical instrumentation, people often hold valuable skills for industry positions. This includes knowledge of optics, software development and project management, and adding the mastery of version control and software management tools can increase one's hireability. 

In summary, version control is the only way to manage software to let you keep a history of all changes, manage diverging aspects of a project and enable efficient collaboration on a project. Every single research project, no matter how small, would benefit from these aspects, and we purposefully include single-person projects here: after all, working by yourself also means collaborating with your past and future self. Who has not come back to a project they have not touched in a couple of months and started scratching their heads as to why it looked the way it did? Testbed projects are very much not like that and usually involve several people working on it at the same time, so collaboration management on the software side becomes key. With different strings of the code needing to be developed in parallel, it becomes almost impossible to develop without using version control, unless the team is willing to take a proliferation of different codes bases into account. This can leave for a very messy consolidation later on, or, in the worst case, loss of work that is kept uniquely on one person's laptop, especially if they end up leaving the group - which is unavoidable in the case of a postdoc or a PhD student.

\subsection{Version control for astronomers}
\label{subsec:version-control-for-astronomers}

\subsubsection{Teaching goals}
\label{subsubsec:teaching-goals}

With the main motivations to use version control in astronomical research, and in particular within testbed teams, listed above, we identified the need to isolate a specific set of skills that are required for a researcher to work with and contribute to a team project using version control. There exists a large number of online training courses and tutorials, paid and free, in-person workshops and classes to learn scientific programming with various programming languages, catering specifically to the scientific community. There are even offers directed especially at astronomers through dedicated summer schools and conference events\cite{CodeAstro,escape,lsst,TheCarpentriesOverall,TheSoftwareCarpentryOverall}. However, we did not identify a comparable offer to engage with version control. Tendentially, exposure to VCS happens as a side note during workshops focusing more on data analysis and scientific computing, and it is rarely given the full attention beyond a general introduction of an hour or two. There is an abundance of general online tutorials on the topic, so much that it is sometimes hard to identify where to start. And there are some git and GitHub teaching resources online aimed at scientists specifically, but most of them do not show a significant difference to generic git tutorials, are geared very much towards data scientists or are sold commercially. There are also some resources about lessons learned on integrating git and GitHub as learning objectives in courses for statistics and data science\cite{Beckman2020}, and some overview materials in the biology community\cite{Blischak2016,Perez-Riverol2016}.

\begin{figure}
    \centering
   \includegraphics[width = 0.9\textwidth]{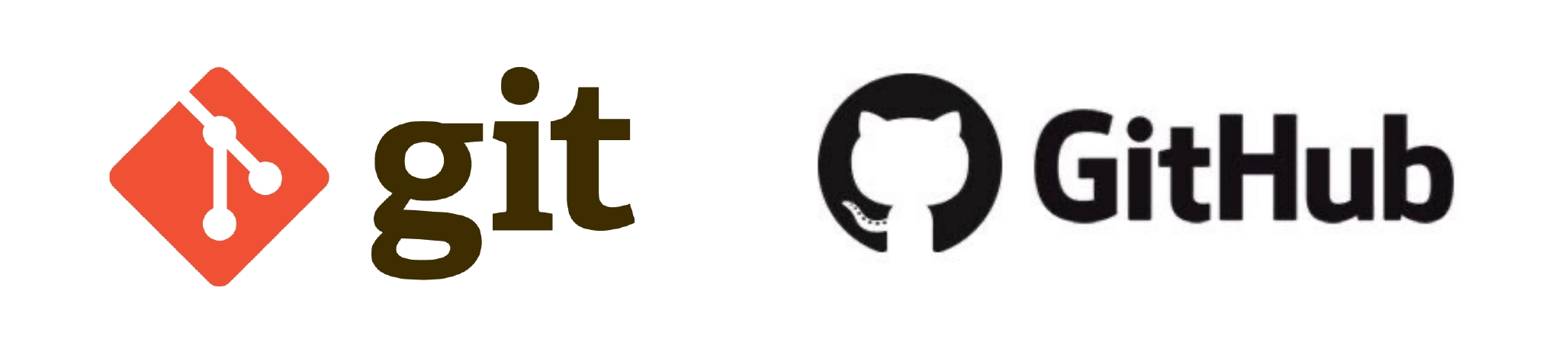}
   \caption[Git and GitHub] 
   {\label{fig:git-and-github} 
    Git is currently the most widely used VCS in research, it is free and open-source, and there is a plethora of online documentation that helps to learn and use it. GitHub is a cloud service based on git that hosts a huge part of today’s open-source software projects. We teach version control jointly with these two tools to provide a solid basis for using VCS in research projects.}
   \end{figure}

This lead to the initiative to design a teaching activity geared particularly toward astronomers. We decided to teach our classes by working jointly with git and GitHub (Fig.~\ref{fig:git-and-github}).
Git ``is a free and open source distributed version control system designed to handle everything from small to very large projects with speed and efficiency.''\cite{git} This means participants of the course did not need to pay for the tool we were teaching, they would find tons of documentation and support online, and they would learn the most widely-used version control system used in the scientific community today. In particular, being a distributed VCS, which makes it far better suited for remote collaborations than traditional centralized VCS\cite{DVCS}. Similar reasons prompted us to work with GitHub: While it is not open-source (it was acquired by Microsoft in 2018), it is free to use, very many projects are already hosted on it and there is no affiliation requirement to be able to sign up (as opposed to institute-internal GitLab access for example). Both tools together facilitate the wide-spread collaborations between researchers and their teams.

The scope and mode of the course was defined by drawing from the experience the co-authors gathered in their respective laboratory teams and optics research groups. In particular, the intent was to address some of the difficulties they witnessed in those groups when it came to on-boarding new members and teaching colleagues how to work with version control. The course goals were developed in such a way that the focus would be the \textit{usage of version control in research projects} and not the version control tool itself. This might seem like a subtle difference and in many ways, it is. However, this approach proved to be way more engaging than pushing through each and every single functionality git has to offer. We thus decided to filter out the main course goals to be:
\begin{itemize}
  \item A general motivation to use version control in research projects
  \item Learn how to be a user and contributor first
  \item Then move on to learning how to manage and create new projects.
\end{itemize}
The order of the bulleted points above matters, for several reasons. One, the target group of our workshop were specifically people who have proven time and time and again that they are smart enough to learn complicated concepts. It would be an easy thing for them to pick up a manual or tutorial that taught them all they needed to know about version control. The crucial point here is to actually spark an interest in them to do so, rather than feeding them information they could find anywhere else. By structuring the workshops in a way that made the usage and application of version control tools more obvious in their concrete work, learning the actual concepts would make be easy for them further down the line. This is why most of the lecture part of the courses talks about an extended version of what is laid out in Sec.~\ref{subsec:version-control-standard}. Two, almost all of the tutorials you find online start off by teaching you how to \textit{create} a repository, but let us face it, how often does any of us really type \verb git ~\verb init ~in their terminal? The bulk of time spent working with version control is not spent on creating new repositories but on maintaining them and contributing to them. Starting off with a skill that is used much rarer by comparison shifts the focus to lower-priority things, while we want to keep them on high-priority things like for example managing branches. Especially considering a testbed project, people will be joining to contribute, not to start their own control code and repository if they have never used git before. That being said, creating new repositories is such a crucial part of using git that of course it was covered as well - but later on in the course.

\subsubsection{A new spin on the same content}
\label{subsubsec:new-spin-same-content}

Following the elaborations in the previous sections, it is clear that the technical content of our training was not much different from any entry-level version control tutorial. It covers the general idea behind version control and some motivation to use it, how git works, what the differences between GitHub and GitLab are, and how to perform basic tasks like cloning a repository, creating branches, committing, inspecting the git history, pushing, pulling, and engaging in pull requests. This would be taught initially from the perspective of a user and contributor, later shifting to the standpoint of a maintainer, curator and creator. The focus here would be on the various workflows one can follow when working alone, in small groups or in teams.

The crucial point we enforced however was a reprioritization of the methods used during the teaching activities, which we captured in Table ~\ref{tab:priorities}. Here, we list more commonly used approaches to teaching version control versus our personal adaptations, as well as the rationale behind the choices.
\begin{table}[h!]
\begin{tabular}{@{}lll@{}}
\toprule
   & Common approach                                          & Adapted approach                                   \\ \midrule
1. & Teach git through the command line interface             & Teach git through a self-standing interactive GUI  \\
2. & Teach git on code examples                               & Teach git on text files                            \\
3. & First lesson: setting up a git user profile and git init & First lesson: create branch, commit, merge         \\
4. & Use individual practice examples                         & Use collaborative practice examples                \\
5. & Teach git first, then introduce GitHub                   & Introduce git and GitHub at the same time \\ \bottomrule
\end{tabular}
\caption[Table of priorities]
{\label{tab:priorities} 
   Major differences between most openly available version control tutorials found online, and the training concept we present in this paper.}
\end{table}

\begin{figure}
    \centering
   \includegraphics[width = 0.9\textwidth]{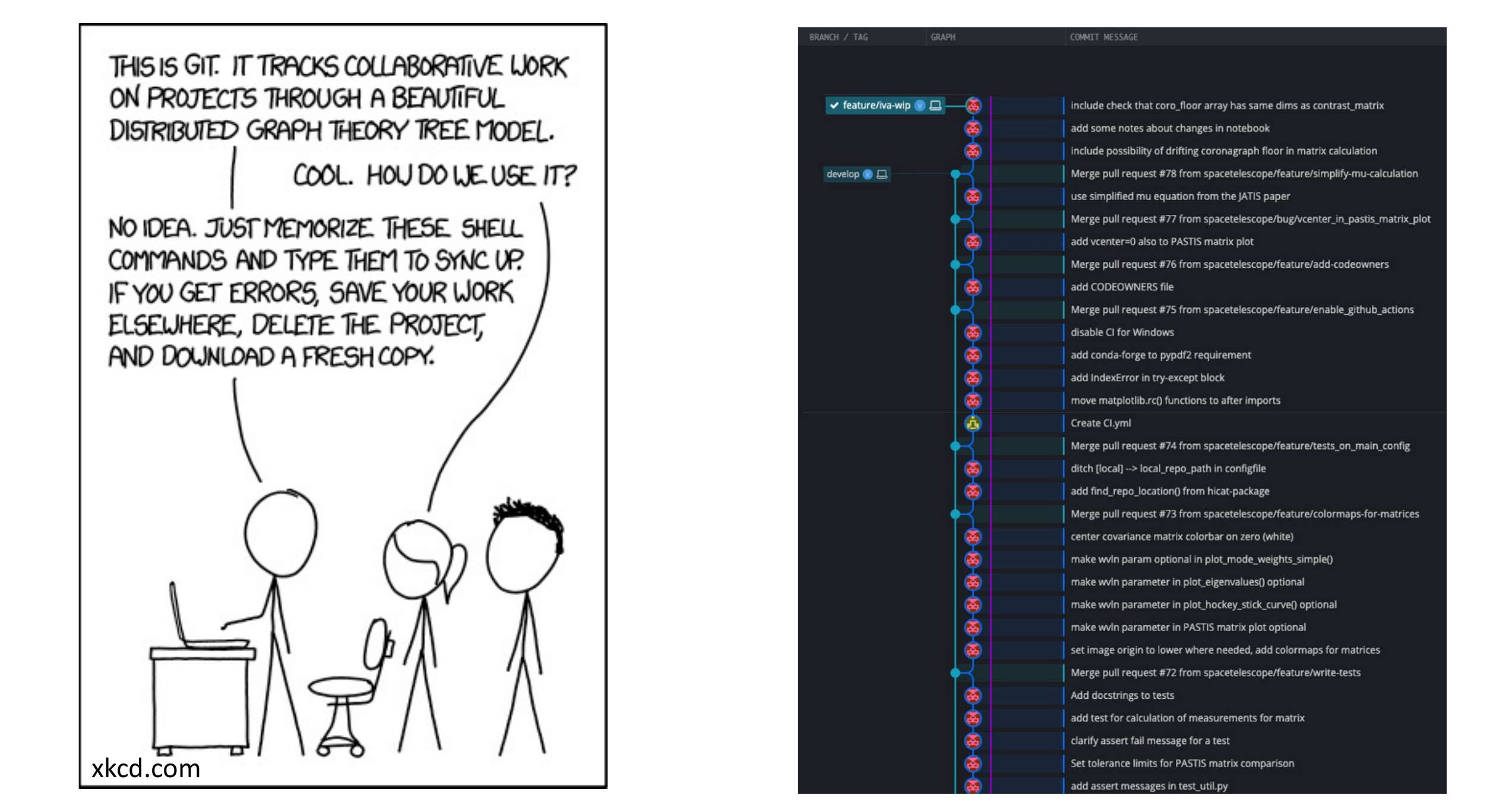}
   \caption[xkcd shell commands] 
   {\label{fig:xkcd-shell-commands} 
    \textit{Left:} While the git CLI is an extremely powerful tool, it also has the major shortcoming of providing little instructional value in teaching git to VCS newcomers. \textit{Right:} Git GUIs are the ideal learning tool because they provide direct feedback of one's actions through a visually rendered history tree like shown here. This example shows a very linear bit of the git history as displayed with GitKraken.}
   \end{figure}

The first major point that stands out in our version control courses is the fact that we never once touch the git command line interface (CLI)\footnote{We do give a brief demo for the sake of completeness, but only \textit{after} the participants had already be using a GUI to perform tasks with git and followed their actions in the visual representation of the git tree.}. Make no mistake: we do not contest its utility or power, but we insist that the basic workings of git fail to be conveyed purposefully by just giving people commands to type in their terminal window (see also Fig.~\ref{fig:xkcd-shell-commands}, left) - it has very limited pedagogical value.
This is in fact an aspect of teaching git that is widely recognized, as exemplified by the following quote by software developer Marco Chiappetta, found on his blog where he talks about how to use git for working on a team:
\begin{quote}
    ``When I had to learn git I started reading lots of articles and asked for help to friends of mine who were more experienced in versioning. They all had the worst approach: they started teaching me how to use the git CLI.''\cite{Chiappetta-quote-2019}
\end{quote}
Clearly, this issue resonates with many people and it is often easier to digest new information if it is accompanied by a visual representation. And yet, even Marco Chiappetta proceeds to provide an introduction to git through the CLI in that very same article! A graphical user interface (GUI) for git overcomes this problem as it unpacks git commands into buttons, labels and dashboards, and most importantly, in most cases it shows a beautifully rendered representation of the git history tree. The GUI we chose for our teaching activities is GitKraken\cite{gitkraken}. We investigated some other GUIs that were candidates for our trainings, like SourceTree, TortoiseGit and SmartGit\footnote{A list of git GUIs can be found here: \url{https://git-scm.com/downloads/guis}}, but GitKraken was the only one we found to satisfy all of the below requirements:
\begin{itemize}
  \item It runs on \textbf{all three major operating systems} (Windows, MacOS, Linux), which means we do not have to change our trainings as a function of OS the participants use.
  \item It is \textbf{self-standing}, meaning it is separated from an editor, IDE (integrated development environment) or file browser.
  \item It includes a \textbf{well rendered graphical representation of the git tree}, which makes it easier to understand what is going on at any given time.
  \item It requires \textbf{no usage of the command line interface at all}, but it is compatible with using it in parallel.
  \item It is \textbf{free of cost} in its basic version, and the Pro version (required to work with private repositories) is free for students and people with an affiliation to an educational institution\footnote{standing as of July 2022}.
  \item It has a good (if not excellent) \textbf{software production quality}, it is \textbf{actively maintained} and it has an extensive online documentation, including their own video tutorials.
\end{itemize}

We especially insisted on using a self-standing client over an integrated one like in the VS Code IDE because one major difficulty we found in teaching version control to newcomers is the natural entanglement of programming with versioning. By having to actively switch to an editor when editing your files, and purposefully opening a git GUI when you are about to perform version control operations, it enhances the point that \textbf{version control has a priori nothing to do with writing code}. It was designed to be a content tracker\cite{Stopak2020} but of course it is inherently optimized for its originally intended use, tracking software changes. To bring this point home, none of our version control trainings involve the use of a programming language, which is noted in point 2 of Table \ref{tab:priorities}. By using simple text files in the interactive exercises and examples\footnote{This is an approach we learned and adapted from the excellent git intro tutorial by The Carpentry\cite{carpentry-git-novice}.} people are free to use whatever editor they like and we avoid the temptation to engage in discussions about preferred programming languages.

Point number 3 given in Table \ref{tab:priorities} has been elaborated in Sec.~\ref{subsubsec:teaching-goals} already: the fraction of time spent working on creating (local) repositories is highly over-represented and often highlighted first in many online tutorials, so we decided to flip around the sequence in which we show people the different parts of a git workflow.

\begin{figure}
    \centering
   \includegraphics[width = \textwidth]{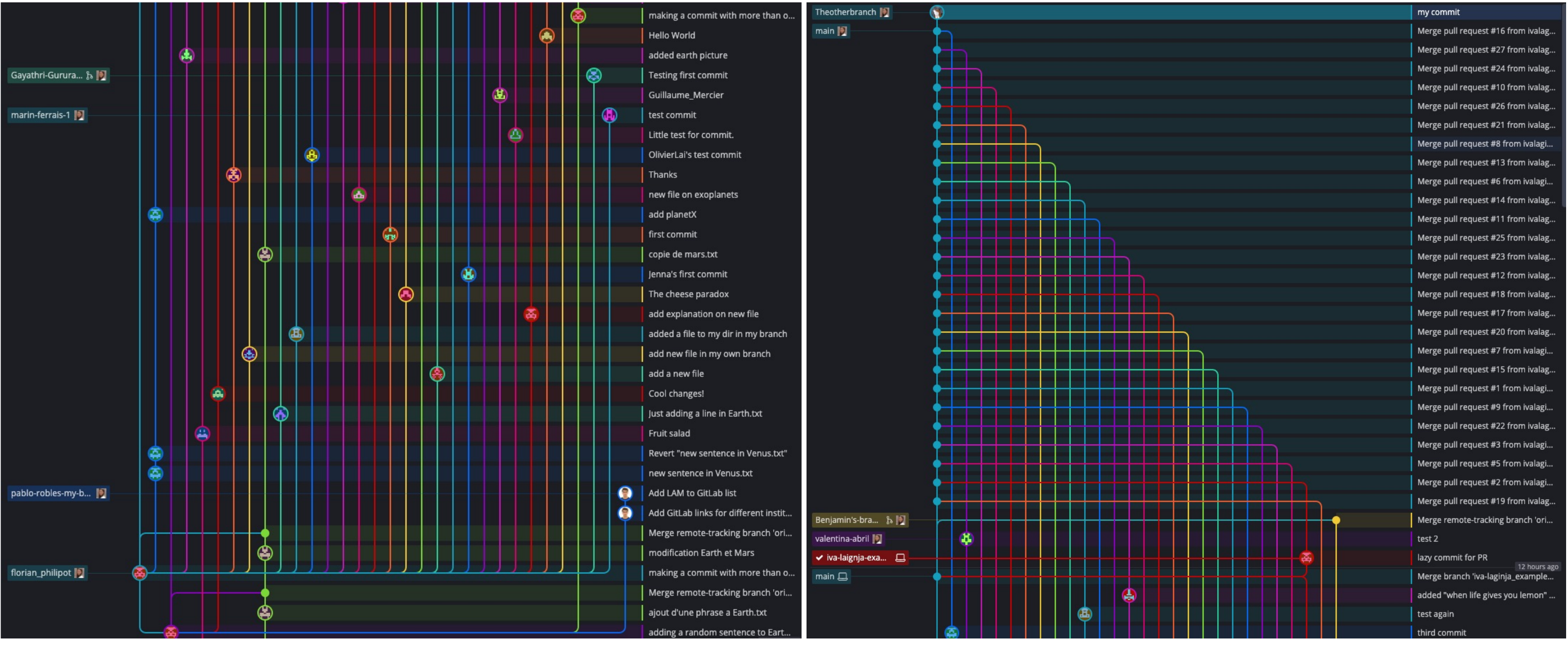}
   \caption[GitKraken example] 
   {\label{fig:gitkraken-example} 
    Hands-on practice examples build the bulk of the git trainings we propose. Through carefully prepared training repositories, the course participants get to learn the basics of git in a highly collaborative way that yields git history trees like shown here. Course attendants learn how to edit content on the same repository by working on separate branches (left) and ultimately how to consolidate all these branches into a single production branch through GitHub pull requests (right).}
   \end{figure}

No version control tutorial would be complete without providing examples and inviting the participants to work through exercises, as indicated by point 4 in Table \ref{tab:priorities}, and our trainings are no different. However, since one of our declared goals is to first teach people how to contribute to collaborative projects, instead of letting everybody create their own repository and practice on there, we immediately walk through an exercise in which all course participants have to contribute to the same repository. In a first step, this is made easy by introducing changes only by creating new files, which avoids merge conflicts. In a later example though, participants are lead to make changes that will purposefully introduce merge conflicts when they open a pull request on GitHub. Here, they are taken through the step-by-step process of resolving them through the GitKraken conflict resolution tool and they review each other's pull requests. By carefully preparing the training repositories and examples, this lead to colorful-looking git trees in the training repository as shown in Fig.~\ref{fig:gitkraken-example} with the result that everybody worked through the same training exercises but in a highly collaborative fashion.
Adpoting the acquired skills to a simpler single-person workflow like the example shown in Fig.~\ref{fig:xkcd-shell-commands}, right, is then just a matter of a new context.

The final point in Table \ref{tab:priorities}, point 5, touches upon the relationship between git and GitHub during training activities. We regard that focusing only on git first, without introducing the concept of a remote repository, is a way of working that no researcher would really ever be confronted with. Even single-user private repositories would be hosted on a remote at some point, so we bring in the joint use of git and GitHub as early as in our very first example of our trainings.

The above account of version control trainings stems from some problems we observed in our day-to-day work, like having to face the git CLI as a newcomer to git, struggling to interpret the history tree and mixing up concepts from programming and from version control. We have found a solution that has served us well in a series of trainings we offered in 2021 and 2022, which we talk about more in the following section. One remaining point to conclude is that while we optimized our trainings for pedagogical efficiency (see Table.~\ref{tab:priorities}), the acquired skills and methods are easily portable to other tools of the user's liking. After being exposed to git and learning about it with a GUI, people can still choose to become a CLI-only user or perform coding and version control from within the same tool. What we argue is that this direction is much easier than the other way around (learn with the CLI if you actually prefer working with GUIs in the end). Likewise, most remote hosting services are based on the same principles, which means that learning how to use them on the most openly available platform (i.e., GitHub) is still very useful to people who then move on to working with something else (e.g., GitLab). The whole concept presented herein aims primarily to maintain a pedagogical narrative.

\subsection{Trainings held so far}

The initial idea for version control courses for astronomers arose in late 2020 and early 2021, which was in the middle of some of the more restricting Covid-19 lockdown periods. As a result, the git trainings created following the principles from the previous sections were designed as a fully remote class held over a video conferencing tool with a screen-share capacity. To keep screen fatigue to a reasonable minimum for a remote work day, we split the git training into two separate sessions, lasting about four hours each, scheduled on two separate days within a few weeks of each other. The first one is titled ``Git for Astronomers Intro'' and the second one``Advanced Git for Astronomers''. The attendance of both modules would give a participant exposure to the full training content and exercises.

Each tutorial is held in such a way that one main instructor is presenting the material and taking the participants through the exercises while a second instructor is available on the group chat to answer questions, bring questions to the attention of the presenting instructor and help out with smaller issues that arise during the training. The introduction class requires the instructors to prepare a training repository on GitHub while the advanced class requires the preparation of three such repositories. Creating them ahead of time with a specific branch and file structure allows for the right merge conflicts to be triggered at the right time of the course. Once set up, these training repositories can easily be used as templates for future trainings.

A first batch of trainings was held remotely in the spring of 2021 while another round of trainings was offered in hybrid mode (class held in-person with possible remote attendance over video and screen share) in early 2022, see Table \ref{tab:trainings-held} for a full list.
\begin{table}[h!]
\centering
\begin{tabular}{cl}
\hline
\textbf{Date (y/m/d)} & \textbf{Course}      \\ \hline
2021 03 17   & Git for Astronomers Intro    \\
2021 04 14   & Git for Astronomers Intro    \\
2021 05 05   & Advanced Git for Astronomers \\
2021 05 19   & Advanced Git for Astronomers \\
2022 04 13   & Git for Astronomers Intro    \\
2022 04 27   & Advanced Git for Astronomers \\ \hline
\end{tabular}
\caption[Table of courses held]
{\label{tab:trainings-held} 
   A list of version control trainings based on our methods held as of July 2022.}
\end{table}
The total number of individual participants across all courses was roughly 80 across all introductory sessions and about 55 across the advanced sessions, where most but not all attendants of the advanced course had also attended the intro class. The attendants included interns, graduate students, postdocs, permanent staff and faculty as well as engineers from all fields in astronomy, including very few participants from other scientific fields (e.g., biophysics). They were affiliated with at least seven different institutions in four different countries (France, Netherlands, Spain, Italy). The feedback was highly positive throughout, especially regarding the alternation between theoretical explanations, practical demonstrations and hands-on exercises, and the exchange between the instructors and attendants.

After the series of trainings described above, the course materials have matured enough to provide a solid basis for an introduction to VCS while also being easily adapted to any specific needs of a group or institute. Further trainings are currently not planned but the authors intend to identify avenues to put the course materials and strategy to good use, for example through online materials, conference workshops or dedicated research group activities.

\section{Summary and conclusions}
\label{sec:summary}

We have presented the broad scope of the CAOTIC project which aims to provide a platform for the astronomical testbed community to connect and exchange beyond the classical pathways of academic publications. The core of the project is currently built by a website that assembles basic information about testbed teams around the world, and their work. The main goal of the project is to identify common aspects of working on HCI and AO testbeds that traditionally get less attention than the actual scientific results, such as hardware handling, project management and software best practices.

As part of this process, we identified the usage of version control systems as a crucial aspect of concrete laboratory work. While we claim that any research project would greatly benefit from engaging with such tools, testbed activities in particular can optimize their work by embracing them. Every single testbed requires the development of a software infrastructure and this is usually done in a highly collaborative manner within a team, but also with its external collaborators. Nevertheless, the motivation for using version control solutions like git are not always recognized or the hurdle to start using it are perceived as too complicated or time-demanding. This led us to conclude that the astronomical research community and in particular researchers working in testbed teams lack appropriate training opportunities to overcome these entry-level barriers.

To change this, we identified some key points that seem to constitute the main difficulties in moving a team to use version control for their projects. We designed a git and GitHub training activity that is adapted around these difficulties and presented its methodology in this paper. We conducted several of such courses in 2021 and 2022, with very positive feedback from the roughly 80 distinct participants. We settled on the use of git and GitHub with the GitKraken GUI, a combination which met the teaching requirements we deducted from the observed difficulties we intended to overcome. While we consider this setup to be the most effective in a pedagogical sense, this does not mean we believe these tools to be the most effective VCS tools for every project. The teaching program we built with our choice of tools makes them easily substitutable with other tools that might be used preferentially by any given user, team or institute, or with new tools that gain relevance in the future.

We would like to note that one of the main observations the authors made in their respective research groups is that in spite of all well-intentioned presentations, provision of tools and demonstrations, there is a tendency on most teams to drop good practices, which includes the use of version control, unless there is at least some level of enforcement. This could either be a top-down decision by the principle investigator (PI), but is usually more efficient if the push comes from within the team, through encouragement and support between the team members. By consequence, it is really the continuous training of and exchange between junior-level researchers that will bring about the changes the CAOTIC project aims to support. With the findings presented in this paper, we hope to incentivize the community to engage in this effort.

\acknowledgments 
I.L. and P.R. would like to thank Mehdi Kourdourli, Alexis Lau and Élodie Choquet for valuable feedback in the early stages of the development of the workshops. I.L. and P.R. would also like to thank Laurent Mugnier for extensive discussions about the core needs from version control in research. I.L. acknowledges the support by a postdoctoral grant issued by the Centre National d'Études Spatiales (CNES) in France.

\section*{CONFLICT OF INTEREST}
The authors declare no conflict of interest and no author holds a commercial or non-commercial affiliation with GitKraken, Resurgens Technology Partners, Axosoft, GitHub or Microsoft.

\bibliography{references}
\bibliographystyle{spiebib}

\end{document}